\documentclass[referee]{aa}

\begin{document}

\title{Poynting-Robertson effect and capture of grains
in exterior resonances with planets}

\author{J.~Kla\v{c}ka and P.~P\'{a}stor}

\institute{Department of Astronomy, Physics of the Earth, and Meteorology, \\
   Faculty of Mathematics, Physics and Informatics,
   Comenius University, \\
   Mlynsk\'{a} dolina, 842~48 Bratislava, Slovak Republic \\
   e-mail: klacka@fmph.uniba.sk; pavol.pastor@fmph.uniba.sk}

\date{}

\abstract{
Spherical meteoroids orbiting the Sun experience orbital decay due to the
Poynting-Robertson effect and can become trapped in commensurability
resonances with a planet. Detail conditions for trapping in the planar circular
restricted three-body problem with action of solar electromagnetic radiation
on spherical grain are presented. The results are given in terms of the
radiation presure factor $\beta$.

In general, the greater value of $\beta$, the shorter capture time in a
resonance. Captures of particles may be even several times longer than it is
presented in literature. Analytical formula for maximum capture times is given.

Analytical theory for minimal capture eccentricities is presented. The obtained
analytical results differ from those presented in the literature. However, the
known analytical results are not very consistent with our detail numerical
computational results for $\beta-$values 0.01 and 0.05 and for the planet
Earth. Minimum eccentricities for captures into resonances are also smaller,
even in order of magnitude, than the published values.

The presented results enable comparison with calculations for nonspherical
particles.

}

\authorrunning{J.~Kla\v{c}ka and P.~P\'{a}stor}
\titlerunning{Capture of grains in exterior resonances with planets}
\maketitle

\section{Introduction}
Physics of resonances with Solar System planets is discussed mainly
during the last three decades. The orbital evolution of meteoroids near
resonances with planets was also investigated. Besides gravitational forces
the effect of solar electromagnetic radiation in the form of the
Poynting-Robertson (P-R) effect is usually taken into account (Robertson 1937,
Kla\v{c}ka 2004, 2008a, 2008b), see e. g., Jackson and Zook (1992),
Weidenschilling and Jackson (1993), Marzari and Vanzani (1994),
\v{S}idlichovsk\'{y} and Nesvorn\'{y} (1994), Liou and Zook (1995),
Liou {\it et al.} (1995).

The aim of our contribution is detail analytical reconsideration of the
influence of the P-R effect on a meteoroid which can become trapped in
commensurability resonances with a planet. As a motivation we can introduce
that Weidenschilling and Jackson (1993) yields for theoretical minimal value
of eccentricity at the beginning of capture the value 0.3506 for dust particle
with $\beta =$ 0.01 in resonance 2:1 with the Earth, while our simulations
show that real values may be even less than 0.1. Detail conditions for trapping
in the planar circular restricted three-body problem with action of solar
electromagnetic radiation on a spherical grain (P-R effect holds) are
presented. We concentrate on real values of orbital elements -- width of
semimajor axis in resonance, minimal values of eccentricities at the
beginning of capture, and, capture time. We use for calculating osculating
orbital elements simultaneously solar gravity and dominant part of solar
radiation pressure-term (central acceleration have form
$-GM_{\odot}(1-\beta)\vec{e}_{R}/r^{2}$). Results presented in this
contribution may serve for comparison of the results obtained for spherical
and nonspherical dust particles (meteoroids), since P-R effect holds only if a
special condition is fulfilled (see Eq. (9) in Kla\v{c}ka 2008a;
compare Eq. (73) in Kla\v{c}ka 2008b) and real nonspherical particles
do not fulfill the condition.

\section{Model -- Equation of motion}

All our numerical simulations are based on equation of motion written in
the form (see Eq. (72) in Kla\v{c}ka 2008b for more details)
\begin{eqnarray}\label{1}
\frac{d \vec{v}}{d t} &=& - ~ \frac{G M_{\odot}}{r^{2}} ~\vec{e}_{R}
      -~ G m_{P} ~ \left \{
      \frac{\vec{r} ~-~ \vec{r}_{P}}{
      | \vec{r} ~-~ \vec{r}_{P} | ^{3}} ~+~
      \frac{\vec{r}_{P}}{| \vec{r}_{P} | ^{3}} \right \}
\nonumber \\
& & + \beta ~\frac{G M_{\odot}}{r^{2}} ~\left \{ \left (
      1 ~-~ \frac{\vec{v} \cdot \vec{e}_{R}}{c} \right ) ~ \vec{e}_{R}
      ~-~ \frac{\vec{v}}{c} \right \} ~,
\nonumber \\
\vec{e}_{R} &\equiv& \vec{r} / |\vec{r}| ~;~~
\beta = 7.682 \times 10^{-4} ~Q'_{pr} ~
      \frac{A' ~\left [ \mbox{m}^{2} \right ]}
      {m ~\left [ \mbox{kg} \right ]} ~,
\end{eqnarray}
where $\vec{r}$ is position vector of the particle (with respect to the Sun)
and $\vec{r}_{P}$ is position vector of the planet which moves in
circular orbit around the Sun; $m_{P}$ is mass of the planet, $M_{\odot}$ is
mass of the Sun, $m$ is mass of the particle/meteoroid,
$A' = \pi ~a^{\prime 2}$, $a^{\prime}$ is radius of the spherical particle,
$Q'_{pr}$ is dimensionless efficiency factor (see, e. g., van de Hulst 1981).
$\beta$ $=$ 5.761 $\times$ 10$^{-5}$ $Q'_{pr}$ $/$
$( \varrho [\mbox{g/cm}^{3}] ~s [\mbox{cm}] )$,
for homogeneous spherical particle in the Solar System, $\varrho$ is
mass density and $s$ is radius of the sphere. Planar motion will be considered.

We use osculating orbital elements which considers solar gravity and dominant
part of solar radiation pressure-term, simultaneously: central acceleration
$-~G M_{\odot} ( 1 - \beta )$ $\vec{e}_{R} / r^{2}$ is used in calculations
of orbital elements and the corresponding quantities will be denoted
with a subscript $\beta$.

\section{Perturbation equations of celestial mechanics}

Rewriting Eq. (1) into the form
\begin{eqnarray}\label{2}
\frac{d \vec{v}}{d t} = &-& ~ \frac{G M_{\odot}
      \left ( 1 - \beta \right )}{r^{2}} ~\vec{e}_{R} -~
      G m_{P} ~ \left \{
      \frac{\vec{r} ~-~ \vec{r}_{P}}{
      | \vec{r} ~-~ \vec{r}_{P} | ^{3}} ~+~
      \frac{\vec{r}_{P}}{| \vec{r}_{P} | ^{3}} \right \}
\nonumber \\
&-&   \beta ~\frac{G M_{\odot}}{r^{2}} ~\left (
      \frac{\vec{v} \cdot \vec{e}_{R}}{c} ~ \vec{e}_{R} ~+~
      \frac{\vec{v}}{c} \right ) ~,
\end{eqnarray}
we can immediately write for perturbation acceleration to Keplerian motion
\begin{eqnarray}\label{3}
\vec{F}_{\beta} &=& \left ( \vec{F}_{\beta} \right ) _{G} ~+~
      \left ( \vec{F}_{\beta} \right ) _{PR} ~,
\nonumber \\
\left ( \vec{F}_{\beta} \right ) _{G} &=& -~ G m_{P} ~ \left \{
      \frac{\vec{r} ~-~ \vec{r}_{P}}{
      | \vec{r} ~-~ \vec{r}_{P} | ^{3}} ~+~
      \frac{\vec{r}_{P}}{| \vec{r}_{P} | ^{3}} \right \} ~,
\nonumber \\
\left ( \vec{F}_{\beta} \right ) _{PR} &=&
      -~ \beta ~\frac{G M_{\odot}}{r^{2}} ~\left (
      \frac{\vec{v} \cdot \vec{e}_{R}}{c} ~ \vec{e}_{R} ~+~
      \frac{\vec{v}}{c} \right ) ~.
\end{eqnarray}

Perturbation equations of celestial mechanics yield for osculating orbital
elements ($a_{\beta}$ -- semimajor axis; $e_{\beta}$ -- eccentricity;
$i_{\beta}$ -- inclination (of the orbital plane to the reference frame);
$\Omega_{\beta}$ -- longitude of the ascending node; $\omega_{\beta}$ --
argument of pericenter; $\Theta_{\beta}$ is the position angle of the particle
on the orbit, when measured from the ascending node in the direction of the
particle's motion, $\Theta_{\beta} = \omega_{\beta} + f_{\beta}$):
\begin{eqnarray}\label{4}
\frac{d a_{\beta}}{d t} &=& \frac{2~a_{\beta}}{1~-~e_{\beta}^{2}} ~
        \sqrt{\frac{p_{\beta}}{\mu \left ( 1 ~-~ \beta \right )}} ~
        \left \{
        F_{\beta ~R} ~e_{\beta}~ \sin f_{\beta} +
        F_{\beta ~T} \left ( 1~+~e_{\beta}~ \cos f_{\beta} \right ) \right \} ~,
\nonumber \\
\frac{d e_{\beta}}{d t} &=&
        \sqrt{\frac{p_{\beta}}{\mu \left ( 1 ~-~ \beta \right )}} ~ \left \{
        F_{\beta ~R} ~ \sin f_{\beta} +
        F_{\beta ~T} \left [ \cos f_{\beta} ~+~
        \frac{e_{\beta} +	\cos f_{\beta}}{1 + e_{\beta} \cos f_{\beta}}
        \right ] \right \} ~,
\nonumber \\
\frac{d i_{\beta}}{d t} &=&
        \frac{r}{\sqrt{\mu \left ( 1 ~-~ \beta \right ) p_{\beta}}} ~
        F_{\beta ~N} ~ \cos \Theta_{\beta} ~,
\nonumber \\
\frac{d \Omega_{\beta}}{d t} &=&
        \frac{r}{\sqrt{\mu \left ( 1 ~-~ \beta \right ) p_{\beta}}} ~
        F_{\beta ~N} ~ \frac{\sin \Theta_{\beta}}{\sin i_{\beta}} ~,
\nonumber \\
\frac{d \omega_{\beta}}{d t} &=& - ~ \frac{1}{e_{\beta}} ~
        \sqrt{\frac{p_{\beta}}{\mu \left ( 1 ~-~ \beta \right )}} ~ \left \{
        F_{\beta ~R} \cos f_{\beta} - F_{\beta ~T}
        \frac{2 + e_{\beta} \cos f_{\beta}}{1 + e_{\beta} \cos f_{\beta}}
        \sin f_{\beta} \right \}
\nonumber \\
& & - ~ \frac{r}{\sqrt{\mu \left ( 1 ~-~ \beta \right ) p_{\beta}}} ~
        F_{\beta ~N} ~ \frac{\sin \Theta_{\beta}}{\sin i_{\beta}} ~
        \cos i_{\beta} ~,
\nonumber \\
\frac{d \Theta_{\beta}}{d t} &=&
        \frac{\sqrt{\mu \left ( 1 ~-~ \beta \right ) p_{\beta}}}{r^{2}} ~-~
        \frac{r}{\sqrt{\mu \left ( 1 ~-~ \beta \right ) p_{\beta}}} ~
        F_{\beta ~N} ~ \frac{\sin \Theta_{\beta}}{\sin i_{\beta}} ~
        \cos i_{\beta} ~,
\end{eqnarray}
where $\mu \equiv G M_{\odot}$ and
$r = p_{\beta} / (1 + e_{\beta} \cos f_{\beta})$;
$F_{\beta ~R}$, $F_{\beta ~T}$ and $F_{\beta ~N}$ are radial, transversal
and normal components of perturbation acceleration.

\subsection{P-R effect and perturbation equations of celestial mechanics}

This subsection follows considerations presented in
Kla\v{c}ka (2004 -- Sec. 6.1), or, in Kla\v{c}ka (1992).
On the basis of Eq. (3), we can immediately write for components of
perturbation acceleration to Keplerian motion:
\begin{equation}\label{5}
\left ( F_{\beta ~R} \right ) _{PR} = -~2~ \beta ~\frac{\mu}{r^{2}} ~
      \frac{v_{\beta~R}}{c} ~,~~
      \left ( F_{\beta ~T} \right ) _{PR} = -~ \beta ~\frac{\mu}{r^{2}} ~
      \frac{v_{\beta~T}}{c} ~,~~
      \left ( F_{\beta ~N} \right ) _{PR} = 0 ~,
\end{equation}
where $\mu \equiv G M_{\odot}$,
$r = p_{\beta} / (1 + e_{\beta} \cos f_{\beta})$
and two-body problem yields
\begin{eqnarray}\label{6}
v_{\beta~R} &=& \sqrt{\mu ~( 1 - \beta ) / p_{\beta}} ~
      e_{\beta} \sin f_{\beta} ~,~
\nonumber \\
v_{\beta~T} &=& \sqrt{\mu ~( 1 - \beta ) / p_{\beta}} ~
      \left ( 1 + e_{\beta} \cos f_{\beta} \right ) ~.
\end{eqnarray}
Inserting Eqs. (5) -- (6) into Eq. (4), one easily obtains
\begin{eqnarray}\label{7}
\left ( \frac{da_{\beta}}{dt} \right ) _{PR} &=& -~\beta ~
      \frac{\mu}{r^{2}} ~ \frac{2 a_{\beta}}{c} ~
      \frac{1 + e_{\beta}^{2} + 2 e_{\beta} \cos f_{\beta} +
      e_{\beta}^{2}  \sin^{2} f_{\beta}}{1~-~e_{\beta}^{2}} ~,
\nonumber \\
\left ( \frac{de_{\beta}}{dt} \right ) _{PR} &=& -~ \beta ~
      \frac{\mu}{r^{2}} ~ \frac{1}{c} ~ \left (
      2 e_{\beta} + e_{\beta}  \sin^{2} f_{\beta} +
      2 \cos f_{\beta} \right ) ~,
\nonumber \\
\left ( \frac{d i_{\beta}}{dt} \right ) _{PR} &=& 0 ~,
\nonumber \\
\left ( \frac{d\Omega_{\beta}}{dt} \right ) _{PR} &=& 0 ~,
\nonumber \\
\left ( \frac{d\omega_{\beta}}{dt} \right ) _{PR} &=& -~ \beta ~
      \frac{\mu}{r^{2}} ~\frac{1}{c} ~ \frac{1}{e_{\beta}} ~ \left (
      2 - e_{\beta} \cos f_{\beta} \right ) \sin f_{\beta} ~,
\nonumber \\
\left ( \frac{d \Theta_{\beta}}{dt} \right ) _{PR}  &=&
      \frac{\sqrt{\mu \left ( 1 - \beta \right ) p_{\beta}}}{r^{2}} ~.
\end{eqnarray}

\subsection{Gravity of a planet and perturbation equations of celestial
mechanics}

As for gravitational acceleration generated by a planet
$\left ( \vec{F}_{\beta} \right ) _{G}$ presented in Eq. (3), we can write
(compare Eqs. (44) in Kla\v{c}ka 2004 for the case $i_{P}$ $=$ $\Omega_{P}$
$=$ $\omega_{P}$ $=$ 0 and Eqs. (57) in Kla\v{c}ka 2004)
\begin{eqnarray}\label{8}
\vec{r}_{P} &=& r_{P} ~\vec{e}_{PR} ~, ~~~ \vec{r} = r ~\vec{e}_{R} ~,
\nonumber \\
\vec{e}_{PR} &=& ( \cos \Theta_{P} , \sin \Theta_{P} , 0 ) ~,
\nonumber \\
\vec{e}_{R} &=& ( \cos \Theta_{\beta} , \sin \Theta_{\beta} , 0 ) ~,
\nonumber \\
\vec{e}_{T} &=& ( -~ \sin \Theta_{\beta} , \cos \Theta_{\beta} , 0 ) ~,
\nonumber \\
\vec{e}_{PR} &=& \cos \left ( \Theta_{P} ~-~ \Theta_{\beta} \right ) ~
      \vec{e}_{R} ~+~
      \sin \left ( \Theta_{P} ~-~ \Theta_{\beta} \right ) ~\vec{e}_{T} ~,
\nonumber \\
\Theta_{P} &=& n_{P} ( t ~-~ t_{0} ) ~+~ \Theta_{P0} ~, ~~~
      n_{P} = \sqrt{G ~ \left ( M_{\odot} ~+~ m_{P} \right )} ~ r_{P}^{-3/2} ~.
\end{eqnarray}
Inserting Eqs. (8) into $\left ( \vec{F}_{\beta} \right ) _{G}$ presented
in Eq. (3), we obtain
\begin{eqnarray}\label{9}
\left ( \vec{F}_{\beta} \right ) _{G} &=&  G m_{P} ~ ( X_{I} ~-~ X_{II} ) ~,
\nonumber \\
X_{I} &=& \frac{\left [ r_{P} ~ \cos \left ( \Theta_{P} ~-~
      \Theta_{\beta} \right ) ~-~
      r \right ] ~\vec{e}_{R} ~+~ r_{P} ~
      \sin \left ( \Theta_{P} ~-~ \Theta_{\beta} \right ) ~
      \vec{e}_{T}}{ \left [ r_{P}^{2} ~+~ r^{2} ~-~ 2 ~r ~ r_{P} ~
      \cos \left ( \Theta_{P} ~-~ \Theta_{\beta} \right ) \right ] ^{3/2}} ~,
\nonumber \\
X_{II} &=& \frac{1}{r_{P}^{2}} ~ \left [
      \cos \left ( \Theta_{P} ~-~ \Theta_{\beta} \right ) ~\vec{e}_{R} ~+~
      \sin \left ( \Theta_{P} ~-~ \Theta_{\beta} \right ) ~\vec{e}_{T}
      \right ] ~.
\end{eqnarray}
In order to obtain $( d a_{\beta} / d t )_{G}$, $( d e_{\beta} / d t )_{G}$,
... $( d \Theta_{\beta} / d t )_{G}$, it is sufficient to put Eqs. (9)
into Eqs. (4).

\section{Exterior mean motion resonances and $d a_{\beta} / d t$}

We will consider only terms proportional to $e_{\beta}^{0}$ and
$e_{\beta}^{1}$, i. e., we will neglect $e_{\beta}^{2}$ and higher orders
in $e_{\beta}$. First of Eqs. (4) can be written in the form
\begin{eqnarray}\label{10}
\frac{d a_{\beta}}{d t} &=& 2~a_{\beta} ~
      \sqrt{\frac{a_{\beta}}{\mu \left ( 1 ~-~ \beta \right )}} ~
      \left \{
      F_{\beta ~R} ~e_{\beta}~ \sin f_{\beta} +
      F_{\beta ~T} \left ( 1~+~e_{\beta}~ \cos f_{\beta} \right ) \right \} ~.
\end{eqnarray}
On the basis of Eqs. (9) we write, then ($\mu_{P} \equiv G m_{P}$):
\begin{eqnarray}\label{11}
\left ( \vec{F}_{\beta ~R} \right ) _{G} &=& \mu_{P} ~ \left \{
      \frac{r_{P} ~ \cos \left ( \Theta_{P} ~-~ \Theta_{\beta} \right ) ~-~
      a_{\beta}}{ X \left ( 0 \right ) } ~-~
      \frac{\cos \left ( \Theta_{P} ~-~ \Theta_{\beta} \right )}{r_{P}^{2}}
      \right \} ~,
\nonumber \\
\left ( \vec{F}_{\beta ~T} \right ) _{G} &=&  \mu_{P} ~ \left \{
      \frac{r_{P} ~ \sin \left ( \Theta_{P} ~-~ \Theta_{\beta} \right ) }
      {X \left ( e_{\beta} \right )} ~-~
      \frac{\sin \left ( \Theta_{P} ~-~ \Theta_{\beta} \right )}{r_{P}^{2}}
      \right \} ~,
\nonumber \\
X \left ( e_{\beta} \right ) &=& \left [
      r_{P}^{2} + a_{\beta}^{2} \left ( 1 - 2 e_{\beta} \cos f_{\beta} \right )
      - 2 a_{\beta} r_{P} \left ( 1 - e_{\beta} \cos f_{\beta} \right )
      \cos \left ( \Theta_{P} - \Theta_{\beta} \right ) \right ] ^{3/2} ~,
\end{eqnarray}
for radial and transversal components of gravitational perturbation
acceleration.

A particle is in exterior mean motion resonance with
a planet when the ratio of their mean motions is approximately
the ratio of two small natural numbers:
$n_{\beta} / n_{P} = j / ( j + q )$ corresponds to $q-$order exterior
resonance. For the first-order resonances we have
$n_{\beta} / n_{P} = j / ( j + 1 )$, i. e., any first-order resonance
may be defined by the ratio $T /T_{P} = ( j + 1 ) / j$.

As for commensurability resonances with a planet moving in the circular
orbit, the third Kepler's law
[$a_{\beta}^{3} n_{\beta}^{2} = \mu ( 1 - \beta )$,
$a_{P}^{3} n_{P}^{2} = \mu ( 1 + m_{P} / M_{\odot} )$]
yields
\begin{equation}\label{12}
\frac{a_{\beta}}{a_{P}} = \left ( 1 ~-~ \beta  \right ) ^{1/3} ~
      \left ( \frac{T}{T_{P}} \right ) ^{2/3} ~
      \left ( 1 ~+~ \frac{m_{P}}{M_{\odot}}  \right ) ^{-1/3}~,
\end{equation}
for the semimajor axis and period of revolution around the Sun
of the particle and the planet (subscript $P$). Defining any resonance
by the ratio $T /T_{P}$, Eq. (12) yields immediately the ratio
$a_{\beta} / a_{P}$.

On the basis of Eqs. (11) we can write for the important resonant terms
\begin{eqnarray}\label{13}
RES &\equiv& \left \{ \left ( \vec{F}_{\beta ~R} \right ) _{G} ~ e_{\beta} ~
      \sin f_{\beta} ~+~
      \left ( \vec{F}_{\beta ~T} \right ) _{G} ~\left ( 1 + e_{\beta} ~
      \cos f_{\beta} \right ) \right \} ~,
\nonumber \\
RES &=& \mu_{P} \left \{
      \frac{r_{P}}{X \left ( 0 \right )} e_{\beta}
      \sin \left ( \Theta_{P} - \omega_{\beta} \right ) -
      \frac{a_{\beta}}{ X \left ( 0 \right ) } e_{\beta}
      \sin f_{\beta} + \frac{r_{P}}{X \left ( e_{\beta} \right )}
      \sin \left ( \Theta_{P} - \Theta_{\beta} \right ) \right \} ~.
\end{eqnarray}
Using the relation
\begin{eqnarray}\label{14}
X \left ( e_{\beta} \right ) &=& X \left ( 0 \right ) ~
      \left \{ 1 ~-~ 3~ e_{\beta} \cos f_{\beta} ~ \frac{ 1 - \varepsilon ~
      \cos \left ( \Theta_{P} - \Theta_{\beta} \right )}{ 1 -
      2 ~ \varepsilon ~\cos \left ( \Theta_{P} - \Theta_{\beta} \right )
      + \varepsilon ^{2}} \right \} ~,
\nonumber \\
      \varepsilon &\equiv& r_{P} / a_{\beta} ~,
\end{eqnarray}
We can Eq. (13) rewrite to the following form
\begin{eqnarray}\label{15}
RES &=& \frac{\mu_{P} ~e_{\beta}}{a_{\beta}^{2} ~
      \left \{ 1 - 2 \varepsilon
      \cos \left ( \Theta_{P} - \Theta_{\beta} \right ) + \varepsilon ^{2}
      \right \} ^{5/2}} ~
      \Biggl \{ - \sin \left ( \Theta_{\beta} - \omega_{\beta} \right )
\nonumber \\
& & + ~ \varepsilon ~ \left [ \frac{7}{2} \sin \left ( \Theta_{P} -
      \omega_{\beta} \right ) - \frac{1}{2} \sin \left ( 2 \Theta_{\beta} -
      \Theta_{P} - \omega_{\beta} \right ) \right ]
\nonumber \\
& & + ~ \varepsilon ^{2} ~ \left [ \frac{7}{4} \sin \left ( \Theta_{\beta} -
      2 \Theta_{P} + \omega_{\beta} \right ) +
      \frac{3}{4} \sin \left ( 3 \Theta_{\beta} - 2 \Theta_{P} -
      \omega_{\beta} \right ) -
      2 \sin\left ( \Theta_{\beta} - \omega_{\beta} \right ) \right ]
\nonumber \\
& & + ~ \varepsilon ^{3} ~
      \sin \left ( \Theta_{P} - \omega_{\beta} \right ) \Biggr \} ~.
\end{eqnarray}
In this expression we neglect terms proportional to
$\sin( \Theta_{\beta}- \Theta_{P})$, because they have not large influence when
we take into account terms important in first-order mean motion resonance. We
use Laplace coefficients $b_{j}$ defined by the relation
\begin{equation}\label{16}
(1 - 2 \varepsilon \cos ( \Theta_{\beta} - \Theta_{P}) +
      \varepsilon^{2})^{-5/2} = \frac{1}{2} \sum_{- \infty}^{+ \infty}
      b_{j} \cos j( \Theta_{\beta} - \Theta_{P}).
\end{equation}
From Eq. (15) using Eq. (16) we get for the first-order exterior resonance
\begin{eqnarray}\label{17}
RES &=& \frac{\mu_{P} ~e_{\beta}}{a_{\beta}^{2}} ~ \frac{1}{2}
      \left [ - \frac{1}{2}~b_{j} +
      \varepsilon \left ( \frac{7}{4}~b_{j+1} - \frac{1}{4}~b_{j-1} \right ) +
      \varepsilon^{2} \left ( -b_{j} - \frac{7}{8}~b_{j+2} +
      \frac{3}{8}~b_{j-2} \right ) \right.
\nonumber \\
& & + \left. \frac{\varepsilon^{3}}{2}~b_{j+1} \right ]
      \sin[(j+1) \Theta_{\beta} - j \Theta_{P} - \omega_{\beta}] ~,
\end{eqnarray}
where we have taken into account only terms with sine argument
$(j+1) \Theta_{\beta} - j \Theta_{P} - \omega_{\beta} = \Gamma$. We define
$C_{j}$ as
\begin{eqnarray}\label{18}
C_{j} \equiv &-& \frac{1}{j+1} \left [ - \frac{1}{2}~b_{j} +
      \varepsilon \left ( \frac{7}{4}~b_{j+1} - \frac{1}{4}~b_{j-1} \right ) +
      \varepsilon^{2} \left ( -b_{j} - \frac{7}{8}~b_{j+2} +
      \frac{3}{8}~b_{j-2} \right ) \right.
\nonumber \\
&+&   \left. \frac{\varepsilon^{3}}{2}~b_{j+1} \right ].
\end{eqnarray}
Using Eqs. (10), (13), (17) and (18) we obtain
\begin{equation}\label{19}
\left ( \frac{d a_{\beta}}{d t} \right ) _{G} = - ~(j+1) ~a_{\beta} ~
      e_{\beta} ~n_{\beta} ~\frac{\mu_{P}}{\mu \left ( 1 - \beta \right )} ~
      C_{j} ~ \sin \Gamma ~.
\end{equation}
For the first-order exterior mean motion resonance Eq. (12) reduces to
\begin{equation}\label{20}
\frac{a_{\beta}}{a_{P}} =
      \left ( 1 ~-~ \beta  \right ) ^{1/3} ~
      \left ( \frac{j ~+~ 1}{j} \right ) ^{2/3} ~
      \left ( 1 ~+~ \frac{m_{P}}{M_{\odot}}  \right ) ^{-1/3}~.
\end{equation}
The values of $C_{j}$ are given in Tab. 1 for particles with
$\beta \in \{0.01,0.05\}$ in the first-order exterior resonance
$j \in \{1,2,3,...,14\}$ with planet Earth in circular orbit.
\begin{table}[!htbp]
\caption
{Values of $C_{j}$ for particle with $\beta=0.01$ and $\beta=0.05$ in
first-order exterior mean motion resonance with the Earth in a circular
orbit.}
\label{T1}
\footnotesize
\vspace{3pt}
\begin{center}
\begin{tabular}{ccc}
\hline
$j$ & $C_{j}(\beta=0.01)$ & $C_{j}(\beta=0.05)$ \\
\hline
1 & 1.31819 & 1.35328 \\
2 & 1.68270 & 1.77804 \\
3 & 2.05585 & 2.23845 \\
4 & 2.43605 & 2.73560 \\
5 & 2.82257 & 3.27183 \\
6 & 3.21514 & 3.85027 \\
7 & 3.61366 & 4.47467 \\
8 & 4.01813 & 5.14933 \\
9 & 4.42856 & 5.87920 \\
10 & 4.84502 & 6.66992 \\
11 & 5.26757 & 7.52797 \\
12 & 5.69630 & 8.46078 \\
13 & 6.13128 & 9.47693 \\
14 & 6.57262 & 10.58637 \\
\hline
\end{tabular}
\end{center}
\end{table}

Neglecting higher orders of eccentricity $e_{\beta}$, the first of Eqs. (7)
yields
\begin{equation}\label{21}
\left ( \frac{da_{\beta}}{dt} \right ) _{PR} = -~2 ~\beta ~
      \frac{\mu}{a_{\beta} ~c} ~
      \left (1 ~+~ 4 e_{\beta} \cos f_{\beta} \right ) ~.
\end{equation}
Eqs. (3) and (4) give
\begin{equation}\label{22}
\frac{d a_{\beta}}{d t} =
      \left ( \frac{d a_{\beta}}{d t} \right ) _{G} ~+~
      \left ( \frac{d a_{\beta}}{d t} \right ) _{PR} ~.
\end{equation}
Exterior resonances can produce long-lived trapping:
$d a_{\beta} / d t =$ 0.
On the basis of Eqs. (19), (21) and (22) we can write
\begin{equation}\label{23}
-~ ( j + 1 ) ~a_{\beta} ~e_{\beta} ~n_{\beta} ~
      \frac{\mu_{P}}{\mu \left ( 1 - \beta \right )} ~C_{j} ~
      \sin \Gamma = 2 ~\beta ~
      \frac{\mu}{a_{\beta} ~c} ~
      \left (1 ~+~ 4 e_{\beta} \cos f_{\beta} \right ) ~.
\end{equation}
Relation $n_{\beta} = n_{P}  ~j / ( j + 1 )$ and the last one of Eqs. (8)
yield $n_{\beta}$ $=$ $\sqrt{\mu}$ $( 1 ~+~ m_{P} / M_{\odot} ) ^{1/2}$
$a_{P}^{-3/2}$ $j / ( j + 1 )$.
Using this result and also Eq. (20) in Eq. (23),
\begin{equation}\label{24}
e_{\beta} = \left \{ 4 \left ( - \cos f_{\beta} \right ) + \frac{1}{2}
      \frac{\mu_{P}}{\mu} \frac{c}{\sqrt{\mu / a_{P}}}
      \frac{\left ( 1 + m_{P} / M_{\odot} \right ) ^{- 1/6}}{\beta
      \left ( 1 - \beta \right ) ^{1/3}}
      \frac{\left ( j + 1 \right ) ^{4/3}}{j^{1/3}}
      C_{j} \left ( - \sin \Gamma \right ) \right \} ^{-1} ~.
\end{equation}
Eq. (24) immediately yields criterion for resonant trapping:
\begin{equation}\label{25}
e_{\beta~min} = \left \{ 4	~+~ \frac{1}{2}~
      \frac{\mu_{P}}{\mu} ~\frac{c}{\sqrt{\mu / a_{P}}} ~
      \frac{\left ( 1 + m_{P} / M_{\odot} \right ) ^{- 1/6}}{\beta
      \left ( 1 - \beta \right ) ^{1/3}} ~
      \frac{\left ( j + 1 \right ) ^{4/3}}{j^{1/3}} ~
      C_{j} \right \} ^{-1} ~.
\end{equation}

\section{Comparison of analytical and numerical results}

We have numerically solved Eq. (1) for various values of $\beta$ and for
various initial conditions. The most important results are presented in Tables
2 and 3 for the case of the first-order exterior resonances with the Earth.
Numerical results are compared with analytical results obtained from
the theory presented in Secs. 2-4 and, also, with the analytical results
presented by Weidenschilling and Jackson (1993).
Our theoretical values of semimajor axis for the exterior resonance
$(j+1)/j$ are obtained from Eq. (20). The values are given in columns
labeled as $a_{\beta}$. The values $a_{\beta~min}$ and $a_{\beta~max}$
are minimal and maximal semimajor axes during the capture in a given resonance.
The width of the resonances in semimajor axis can be calculated from the
relation $a_{\beta~max} - a_{\beta~min}$. Minimal values of eccentricities
for the capture into resonances, found by numerical solution of Eq. (1),
are given in the column $e_{\beta~min~num.}$. Minimal capture eccentricities
$e_{\beta~min~th.}$ were computed from Eq. (25). Minimal capture
eccentricities $e_{\beta~min~th.~WJ}$, presented in Tab. 2, are taken from
Table II in Weidenschilling and Jackson (1993). Maximal capture times
for different first-order resonances are also given. The times and
the minimum and maximum semimajor axis values are obtained for a given
numerical integration of a meteoroid. The extremal values of semimajor axis
occur when the maximal capture time exists, for a given resonance.
The minimum eccentricity values were found independently. There is no
relation between the minimum eccentricity and the set
$\{$maximal capture times, minimum and maximum semimajor axis values$\}$,
as for the presentations in Tables 2 and 3.

\begin{table}[!htbp]
\caption
{Characteristics for the first-order $(j+1)/j$ resonances with Earth for
particle with $\beta=0.01$.}
\label{T2}
\footnotesize
\vspace{3pt}
\begin{center}
\begin{tabular}{cccccccc}
\hline
$j$ & $a_{\beta}$ & $a_{\beta~min}$ & $a_{\beta~max}$ & $e_{\beta~min}$ &
$e_{\beta~min}$ & $e_{\beta~min}$ & maximal \\
 & & & & $_{num.}$ & $_{th.}$ & $_{th.~WJ}$ & capture time \\
 & [ AU ] & [ AU ] & [ AU ] & & & & [ $10^{3}$ years ] \\
\hline
1 & 1.5821 & 1.5761 & 1.5881 & 0.102 & 0.1107 & 0.3506 & 1492 \\
2 & 1.306 & 1.3023 & 1.3097 & 0.036 & 0.0784 & 0.0387 & 736 \\
3 & 1.2074 & 1.2046 & 1.2101 & 0.011 & 0.0564 & 0.0227 & 742 \\
4 & 1.1565 & 1.1544 & 1.1586 & 0.008 & 0.0418 & 0.0148 & 429 \\
5 & 1.1255 & 1.1238 & 1.1272 & 0 & 0.032 & 0.0104 & 268 \\
6 & 1.1045 & 1.1031 & 1.1059 & 0.003 & 0.025 & 0.0077 & 252 \\
7 & 1.0894 & 1.0881 & 1.0907 & 0.001 & 0.0201 & 0.0059 & 227 \\
8 & 1.0781 & 1.077 & 1.079 & 0 & 0.0164 & 0.0046 & 362 \\
9 & 1.0692 & 1.0681 & 1.0701 & 0.001 & 0.0136 & 0.0038 & 167 \\
10 & 1.062 & 1.061 & 1.0626 & 0 & 0.0114 & 0.0031 & 469 \\
11 & 1.0562 & 1.0551 & 1.0566 & 0.002 & 0.0097 & 0.0026 & 75 \\
12 & 1.0513 & 1.0503 & 1.0516 & 0.001 & 0.0084 & 0.0022 & 36 \\
13 & 1.0471 & 1.0459 & 1.0479 & 0.001 & 0.0073 & 0.0019 & 21 \\
14 & 1.0436 & 1.0423 & 1.0444 & 0.001 & 0.0064 & 0.0016 & 13 \\
\hline
\end{tabular}
\end{center}
\end{table}

\begin{table}[!htbp]
\caption
{Characteristics for the first-order $(j+1)/j$ resonances with Earth for
particle with $\beta=0.05$.}
\label{T3}
\footnotesize
\vspace{3pt}
\begin{center}
\begin{tabular}{ccccccc}
\hline
$j$ & $a_{\beta}$ & $a_{\beta~min}$ & $a_{\beta~max}$ & $e_{\beta~min}$ &
$e_{\beta~min}$ & maximal \\
 & & & & $_{num.}$ & $_{th.}$ & capture time \\
 & [ AU ] & [ AU ] & [ AU ] & & & [ $10^{3}$ years ] \\
\hline
1 & 1.5605 & 1.5555 & 1.5656 & 0.196 & 0.1981 & 216 \\
2 & 1.2882 & 1.2851 & 1.2911 & 0.096 & 0.1701 & 168 \\
3 & 1.1909 & 1.1887 & 1.1931 & 0.061 & 0.1422 & 187 \\
4 & 1.1407 & 1.1389 & 1.1426 & 0.045 & 0.1172 & 80 \\
5 & 1.1101 & 1.1088 & 1.1114 & 0.028 & 0.096 & 56 \\
6 & 1.0894 & 1.0879 & 1.091 & 0.018 & 0.0786 & 102 \\
7 & 1.0746 & 1.0735 & 1.0756 & 0.016 & 0.0645 & 62 \\
8 & 1.0633 & 1.0626 & 1.0643 & 0.01 & 0.0531 & 23 \\
9 & 1.0546 & 1.0538 & 1.0552 & 0.007 & 0.044 & 20 \\
10 & 1.0475 & 1.0466 & 1.0481 & 0.005 & 0.0366 & 13 \\
11 & 1.0418 & 1.0408 & 1.0424 & 0.005 & 0.0307 & 6 \\
12 & 1.0369 & 1.036 & 1.0376 & 0.004 & 0.0258 & 3 \\
13 & 1.0328 & 1.0317 & 1.0334 & 0.005 & 0.0218 & 5 \\
14 & 1.0293 & 1.028 & 1.0301 & 0 & 0.0185 & 1 \\
\hline
\end{tabular}
\end{center}
\end{table}

General trend seen from Tables 2 and 3 is that the larger value of $j$ --
type of the first-order exterior resonances with the Earth is $(j+1)/j$,
the smaller width of resonances in semimajor axes and the shorter capture
times. Moreover, the greater value of $\beta$,
i) the smaller resonant width in semimajor axes, and,
ii) the shorter capture time.

\section{Extrapolation of capture times}

We found, from numerical integrations of Eq. (1), that maximal capture time
for a particle in a given mean motion resonance is proportional to the value
of the particle's $\beta$ and semimajor axis $a_{\beta}$ according to
\begin{equation}\label{26}
\tau \sim \frac{a_{\beta}^{2}}{\beta} ~.
\end{equation}
This relation is also in accordance with the expected dependence of secular
time derivation of eccentricity (Liou and Zook 1997; Kla\v{c}ka et al. 2008 --
$\eta$ $\equiv$ 0) on $a_{P}$ and $\beta$. Using Eq. (12)
(with assumption $m_{P} \ll M_{\odot}$) we obtain from Eq. (26)
\begin{equation}\label{27}
\frac{\tau(a_{P1}, \beta_{1})}{\tau(a_{P2}, \beta_{2})} =
\left (\frac{a_{P1}}{a_{P2}} \right )^{2} \frac{\beta_{2}}{\beta_{1}}
\frac{(1 - \beta_{1})^{2/3}}{(1 - \beta_{2})^{2/3}} ~.
\end{equation}
Eq. (27) presents the analytical formula for obtaining maximum capture time
$\tau(a_{P2}, \beta_{2})$ for the particle of $\beta_{2}$ and planetary orbital
radius $a_{P2}$ for a given type of resonance, if maximum capture time
$\tau(a_{P1}, \beta_{1})$ is known for for the particle of $\beta_{1}$ and
planetary orbital radius $a_{P1}$ for the same type of resonance.

Using Eq. (27) we have calculated the values of capture times in Tab. 4
from values of capture times in Tables 2 and 3. We use the
values in Tab. 2 in calculations of the expected values in Tab. 3, and vice
versa.
\begin{table}[!htbp]
\caption
{Extrapolation of capture times presented in Tables 2 and 3. Eq. (27) is used.}
\label{T4}
\footnotesize
\vspace{3pt}
\begin{center}
\begin{tabular}{ccc}
\hline
$j$ & extrapolated & extrapolated \\
 &  values in Tab. 2 & values in Tab. 3 \\
 & [ $10^{3}$ years ] & [ $10^{3}$ years ] \\
\hline
1 & 1110 & 290 \\
2 & 863 & 143 \\
3 & 961 & 144 \\
4 & 411 & 83 \\
5 & 288 & 52 \\
6 & 524 & 49 \\
7 & 319 & 44 \\
8 & 118 & 70 \\
9 & 103 & 32 \\
10 & 67 & 91 \\
11 & 31 & 15 \\
12 & 15 & 7 \\
13 & 26 & 4 \\
14 & 5 & 3 \\
\hline
\end{tabular}
\end{center}
\end{table}

\section{Conclusion}

The planar circular restricted three-body problem with action of solar
electromagnetic radiation on a spherical grain is considered. This
Poynting-Robertson effect is studied numerically and analytically.
We have found minimal values of capture eccentricities into the first-order
exterior mean motion resonances. Unfortunately, our theory for the minimal
values of the capture eccentricities cannot explain the found numerically
values. The numerical values $e_{\beta~min}$ are also smaller than the
values presented in Weidenschilling and Jackson (1993). Thus,
any analytical method explaining minimal values of capture
eccentricities does not exist, for the first-order exterior mean motion
resonances.

Our numerical results offer several resonant characteristics for
$\beta-$values 0.01 and 0.05 for the first-order exterior resonances with the
Earth; capture times for $\beta =$ 0.1 are in order of magnitude shorter than
for $\beta =$ 0.05. We have succeeded in finding greater values of capture
times than it is presented in \v{S}idlichovsk\'{y} and Nesvorn\'{y} (1994).
Eq. (27) presents an analytical formula for obtaining maximum capture time
for the particle of $\beta_{2}$ and planetary orbital radius $a_{P2}$
for a given type of resonance, if maximum capture time is known for
for the particle of $\beta_{1}$ and planetary orbital radius $a_{P1}$
for the same type of resonance.

The presented values can be useful in comparison of resonant evolution
for spherical and nonspherical meteoroids. Our results, based on the
P-R effect, can be used for spherical particles, only. Real nonspherical
particles may exhibit nonzero values for other two components of the
radiation pressure efficiency factor vector $Q'_{2}$ and $Q'_{3}$
(Kla\v{c}ka 2004), and, moreover, values of $\beta-$parameter are time
dependent. Thus, the change of semimajor axis $a_{\beta}$ in orbital
resonance caused by the change of parameter $\beta$ may be relatively large.
Since the resonant width is a decreasing function of $\beta$, one should await
that resonant capture times for nonspherical particles are smaller than the
capture times for spherical grains.

\acknowledgements{}
The paper was supported by the Scientific Grant Agency VEGA (grant No.
2/0016/09).

\end{document}